\documentclass[aps,prb,notitlepage,showpacs,superscriptaddress,groupedaddress,footinbib]{revtex4}
\usepackage[dvips]{graphicx}
\usepackage{subfigure}
\usepackage{epstopdf}
\usepackage{color}
\usepackage{float}
\usepackage{ucs}
\usepackage[utf8x]{inputenc}

\begin{document}


\title{Efficacious calculation of Raman spectra in high pressure hydrogen}

\author{G.J.Ackland$^{\ast, a}$ and I.B.Magdau$^{a}$}
\affiliation{ $^{a}${\em{CSEC and SUPA, University of Edinburgh, Edinburgh, EH9 3JZ;}}}
\thanks{$^\ast$Corresponding author. Email: gjackland@ed.ac.uk
\date{September  2013} }
\begin{abstract}
We present and evaluate an efficient method for simulating Raman
spectra from molecular dynamics (MD) calculations {\it without}
defining normal modes. We apply the method to high pressure hydrogen
in the high-temperature ``Phase IV'': a plastic crystal in which the
conventional picture of fixed phonon eigenmodes breaks down.
  Projecting
trajectories onto in-phase molecular stretches is shown to be many
orders of magnitude faster than polarisability calculations, allowing
statistical averaging at high-temperature. 
 The simulations are extended into
metastable regimes and identify several regimes associated with
symmetry-breaking on different timescales, which are shown to exhibit
features in the Raman spectra at the current experimental limit of
resolvability.  In this paper we have concentrated on the methodology,
a fuller description of the structure of Phase IV hydrogen is given in
a previous paper\cite{Magdau}.

\end{abstract}

\maketitle

\section{Introduction}
Hydrogen is the simplest element, yet its high pressure phase diagram
is not well established.  This is because of the very high
pressures at which interesting behaviour occurs, because the
structure are highly anharmonic and because crystallography is
very challenging.  Samples cannot be made of sufficient size to
perform neutron scattering at the required pressures, while the X-ray
scattering cross-section is weak\cite{akahamaX}.  Consequently, almost
all information about Phase IV comes from spectroscopic
experiments\cite{Eremets,Howie2012} and ab initio calculations
\cite{Pick2007,pick2012,Liu,Magdau}.  To make progress we need to be
able to compare these data directly to enable a joint experiment/theory attack on the problem, as in our previous work in the related first group elements at pressure\cite{MarquesLi,LiLD,MarquesK,LarsK,MarquesNa}.

For a unit cell with $N$ atoms there are exactly 3N phonons
(i.e. frequencies and eigenmodes) at any wavevector $q$.  The phonons
relevant to spectroscopy are those at $\Gamma (q = 0)$.  The
traditional method for spectroscopy simulations is to calculate these
phonons, and to understand how they evolve at high temperature. Once
the phonons are calculated, Infrared or Raman activity is calculated,
thus making the bridge to experiment.

Lattice Dynamics (LD) is the well-established method for phonon
calculations based on second derivatives of the energy function.  With
ab initio methods these can be evaluated either numerically, or by
perturbation theory\cite{mcw,dfptcastep}. 
The phonon spectrum can also be
 calculated by Fourier transformation of the velocity
 autocorrelation function calculated by MD.  
This automatically incorporates anharmonic effects.

Infrared spectra arise from the absorption of energy by a phonon modes
which transitions to an excited level.  Harmonic oscillations of a
particular frequency in the crystal will absorb light at that
frequency, leaving gaps in the transmitted light spectrum, provided
that the eigenmode is IR active.  IR intensity is controlled by the
change in the electric dipole moment occurring in a given oscillation
mode. The intensity of a transition from state m to state n is given
by:
\[ I_{nm} \propto \langle m|\mu_a|n\rangle^2 =\sum_a (\int \Phi^*_m \mu_a \Phi_n)^2; \hspace{1cm}  \mu_a = \mu_a^0 + \sum_k \frac{\partial \mu_a}{\partial \epsilon_{ka}}  \epsilon_{ka} \] with $a$ being one of the Cartesian components and
the dipole moment  $\mu_a$ expressed in the basis of the eigenmodes $ \epsilon_{ka} $. 
With orthogonal states $\Phi_m$, $\Phi_n$, the $\mu_a^0$ term does not contribute, so the only contribution to the intensity comes from the change of dipole moment, which can be calculated from the total energy as a linear response\cite{dfptcastep}:
\[ \frac{\partial \mu_a}{\partial \epsilon_{ka}} = - \frac{\partial^2 V(r)}{\partial A_a\partial \epsilon_{ka}}
= - \frac{\partial F_k}{\partial A_a} 
\]
 where $A_a$ is an external field and $F_k$ are the forces corresponding to each mode $k$.  As
such, the IR activity can be calculated using only first order
derivatives of the forces.

In the classical picture of Raman spectroscopy, the external field
(laser) induces a dipole moment
proportional to the polarisability tensor.  This induced moment
interacts with the field in a second order process: Raman-active
phonon eigenmodes does not themselves induce a dipole moment.  The
intensity depends on third derivatives of the energy\cite{ramanDFPT}:
\[ P_{kab}  \propto \frac{\partial^3E(r)}{\partial A_a \partial A_b \partial \epsilon_{ka}} = \frac{\partial^2F_k}{\partial A_a \partial A_b} \] 
So matrix elements and cross-terms linking
wavefunctions at different $k$-values must be calculated. This
has two important practical effects;  Raman intensity
calculations are {\it much} slower than their IR counterparts, and are
highly sensitive to $k$-point sampling.
The motivation for the present work is to 
find proxies for expensive explicit
polarisability  calculation, so as to devote
computing resource to better sampling of the pressure-temperature space. 
\section{Extracting Phonons from MD}
From MD we generate atomic trajectories, ${x}_{i}(t)$, at finite temperature.  We can expand 
Cartesian components
in terms of the normal modes (ignoring translations).
\[ x_{j}(t) =  X_j+ \sum_{i=4}^{3N}\alpha_{i}(t) {\epsilon}_{ij};  
\hspace{1cm}
{\dot{x}}_{j}(t) =  \sum_{i=4}^{3N}\dot{\alpha}_{i}(t) {\epsilon}_{ij} 
\]
This is simply a linear transformation, the normal mode coefficients
$\alpha_{i}(t)$ are fully determined by the Cartesian positions
$x_{j}(t)$.  If we assume that we are in the harmonic regime,
$\dot{\alpha}_{i}(t) = Im \left [
  a_i\omega_i\exp^{i(\omega_it+\phi_i)} \right]$ with $a_i$,
$\omega_i$ and $\phi_i$ independent of time.  This means we can use
Fourier Transformed MD data to obtain $\omega_i$.  Using the velocity
FT is most convenient because anharmonicity may move the mean
positions at high temperature, because $\langle \alpha_i\rangle \ne
0$, but $\langle \dot{\alpha_i} \rangle= 0 \forall i $.  In the
long-time harmonic limit
 $FT[\dot\alpha_i(t)]$ is simply a delta function at $\omega=\omega_i$, in 
practice we obtain a peak.  

This process uses the LD eigenmodes but not the frequencies: finite temperature phonons are 
calculated from MD, and in the case of strong mixing of LD modes
the Fourier transform may show more than one peak. 
In the harmonic approximation, the
same modes will be Raman/IR active in MD and LD.
The occupied phonon density of states can be found from the MD velocity autocorrelation function:
\[ FT \left [\sum_j \dot{x}_j(t)\dot{x}_j(0) \right ] = FT \left [\sum_i\sum_j 
\dot{\alpha}_i(t)\dot{\alpha}_i(0)e_{ij}^2 \right ] \]
By analogy, the total Raman tensor becomes:
$ FT \left [\sum_{ij} P_{iab} \dot{\alpha}_i(t)\dot{\alpha}_i(0) e_{ij}^2 \right ] $
and the phonon frequency for each mode $i$ comes from the peak  in: $FT \left [ \dot{\alpha}_i(t)\dot{\alpha}_i(0)  \right ] $.
In the harmonic limit, the Raman signal is simply the sum of individual modes. 

Previous methods to obtain high-temperature phonons are based on the idea
that the phonon eigenvectors are temperature independent\cite{Pinsook,Souvatzis}. 
However,  Raman intensities are related to atomic motions, so in the classical 
approximation they can be extracted directly from linear combinations {\it without} explicitly evaluating eigenmodes.

\begin{figure}
\begin{center}
\resizebox*{10cm}{!}{\includegraphics{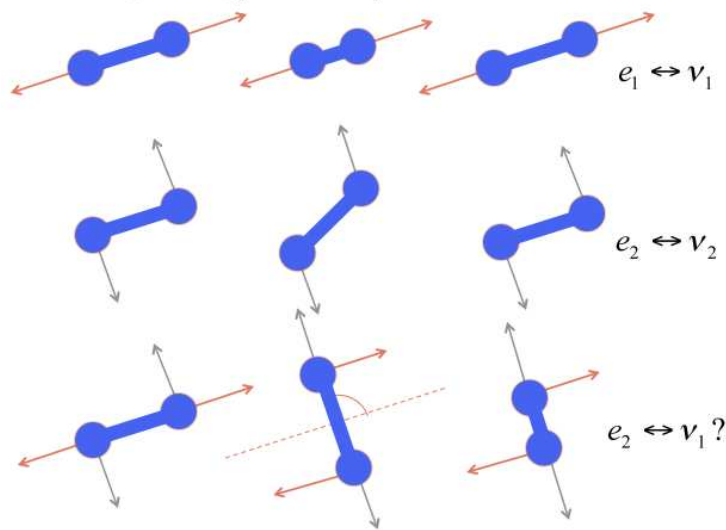}}
\caption{Failure of projection method for rotating molecule.  The upper three images depict a vibron ($\nu_1$),  red arrows showing eigenmode and dumbbell showing associated  molecular deformation.  The central three images show a libron ($\nu_2$, black arrows).  The lower three images show the molecule having rotated through 90 degrees, while the eigenvectors remain fixed in Cartesian coordinates.  The red arrows now correspond to the libron, and the black to the vibron. Thus one cannot use normal modes defined in real space for plastic crystals\label{fig2}}
\end{center}
\end{figure}

For hydrogen the situation is still harder: over the duration of an MD
simulation the molecules can rotate, which can totally change the
nature of an eigenmode defined by a Cartesian eigenvector (see Fig.1); a method is needed which
does not rely on eigenmodes.  The key to this is that
Raman activity in vibrons comes from in-phase molecular vibrations.
For phase III there is only
one such mode: all molecules are in similar environments.  However in
Phase IV the hydrogens in the two layers (B and G, see Fig.2) have very different bond
strengths, and consequently do not couple, giving two Raman modes at
different frequencies.  These are still modes in which all hydrogens
vibrate in phase, however the amplitude of vibration is effectively
zero in one layer or the other. 

This all suggests
that rather than projecting onto eigenmodes, we can obtain Raman intensities by
projecting the MD trajectories directly onto the in-phase molecular vibration.  
This requires us to identify molecules at each step of
the simulation, which can be done simply by taking 
shortest bondlengths in almost every case.  Our Raman-activity proxy is then:
$\alpha_{Raman}(t) = \sum_{j}[{\bf r_j(t) - r_{j_m}(t)}]$
where ${\bf r}_{j_m}(t) $ is the position of the molecular partner atom to $j$, at time $t$.
The spectrum for the vibron modes is extracted from velocity autocorrelation of  $\alpha_{Raman}(t)$: 
\[ FT \left [\ \sum_{j}{\bf v_j(t)\cdot[r_j(t) - r_{j_m}(t)]} \right ]\]
Here we investigate vibrons, but the method is completely general provided
that the Raman-active molecular mode can be identified.
\section{Molecular Dynamics}
Molecular hydrogen comprises two indistinguishable protons (fermions)
and must have an overall antisymmetric wavefunction.  At low pressure
it adopts two forms, ortho- and para-, depending on the nuclear spin
and rotational quantum number J.  In phase I , multiple roton bands
and a ratio of 2:1 between hydrogen and deuterium roton modes is
observed, indicating that these are indeed free rotors and not
harmonic oscillators.  For vibrons, the ratio close to
$\sqrt{2}$:1, showing that these are phonons\cite{footnote}.  Moreover, for
Raman frequencies we should consider the excited state of the
phonon vibration, whose energy corresponds to several thousand Kelvin,
well in excess of the melting point.  

All of these effects
are neglected both in classical MD\cite{footnote2}, and the most commonly used
alternative, path integral MD, which considers the ground state
assuming uncorrelated, distinguishable protons.  Nevertheless, the
equivalence principle suggests that the vibrations in our calculations
should have frequencies corresponding to observation, because the
frequency of a harmonic oscillator transition is independent of its
quantum number.  Although a proper quantum treatment of the proton has
not been done here or elsewhere, it is reasonable to assume that the
main effect will come from the zero point motion and
wavefunction. This in turn will increase the effective displacement of
the protons from their equilibrium positions, and although the
sampling statistics are different\cite{colournoise}, 
to a first approximation it will be equivalent to an increased temperature.

The nature of the phase IV as observed in molecular dynamics is
strongly dependent on finite size effects.  Small simulations with less
than 100 atoms are subject to fluctuations larger than the system
size, leading to spurious effects such as layer reconstructions
fluctuations between B and G or apparent rapid diffusion. All of our
results are taken from simulations of 288 atoms or more.

A stringent test to validate our ansatz and code is provided by
considering a randomised hydrogen-deuterium mixture: the different
molecular masses on HH, HD, DD break the symmetry and mix all the
vibron modes such that they all have some Raman activity.  Figure 2
compares results from our proxy with full 
polarisability calculation\cite{castep,ramanDFPT}.  The perturbation
theory (DFPT)
calculation was done using 1000 processors in 1 week, by contrast the
mode projection took 2 minutes on a single processor.
Agreement is excellent with slightly more mode-mixing in the MD as one
would expect at non-zero temperature.  At these low temperatures
comparison with a single polarisability calculation is reasonable.  At
high temperatures the polarisability calculation would need to be done
at every MD timestep, and averaged, in order to account for anharmonic
temperature effects.

\begin{figure}
\begin{center}
\subfigure[Calculated Raman Spectrum]{
\resizebox*{10cm}{!}{\includegraphics{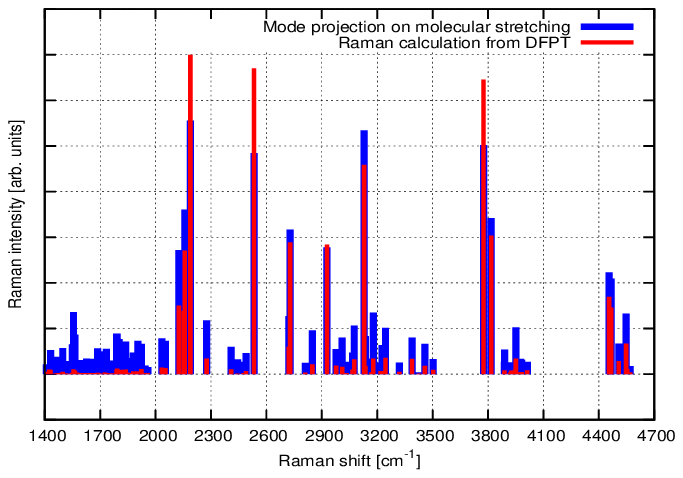}}}
\subfigure[Phase IV hydrogen]{
\resizebox*{3.2cm}{!}{\includegraphics{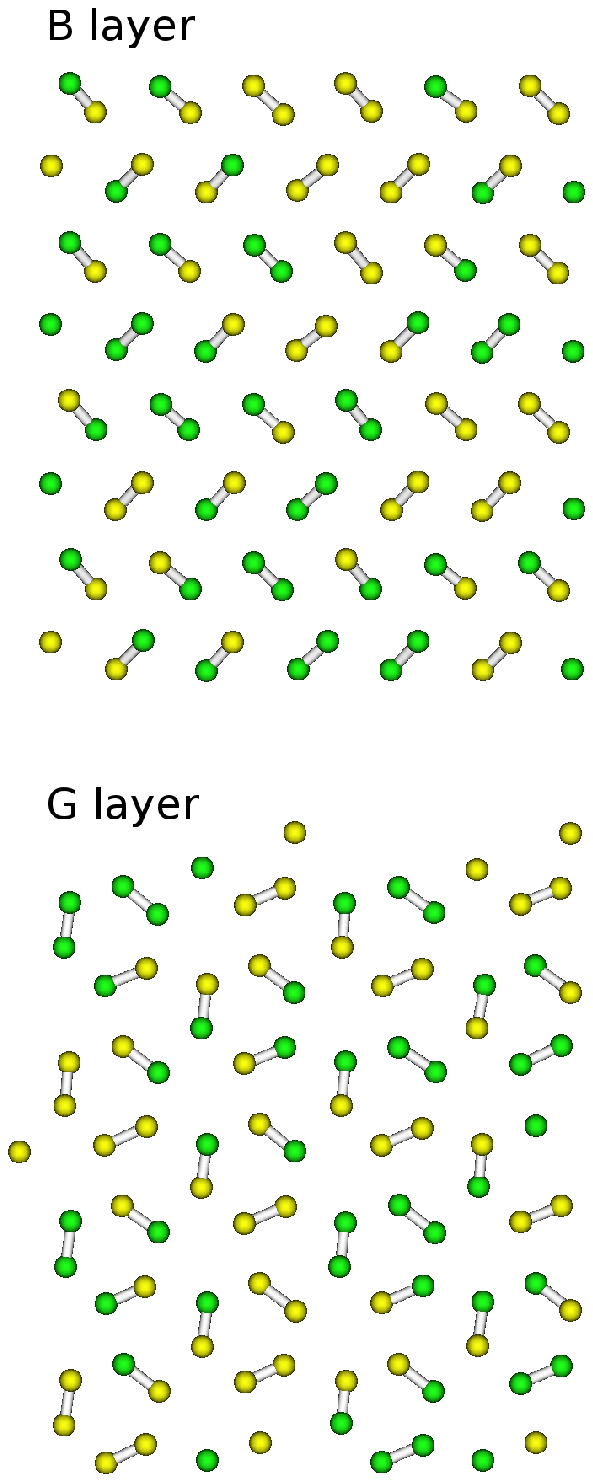}
}}
\caption{Comparison of calculated Raman vibron intensities using polarisability from third-order wavefunction derivatives (red) and projection from molecular dynamics (blue)
for Phase IV with randomly placed H-D isotopes, within the $Pc$ structure\cite{Pick2007,pick2012,Magdau} which has alternating molecular (B) and trimer (G) layers.
\label{fig1}}
\end{center}
\end{figure}

\begin{figure}
\begin{center}
\subfigure[Variation with temperature, 250GPa]{
\resizebox*{5cm}{!}{\includegraphics{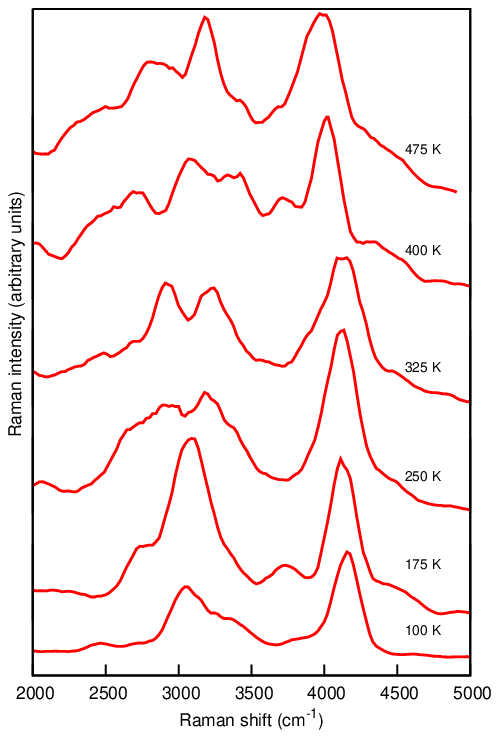}}}
\subfigure[Variation with pressure, 220K.]{
\resizebox*{5cm}{!}{\includegraphics{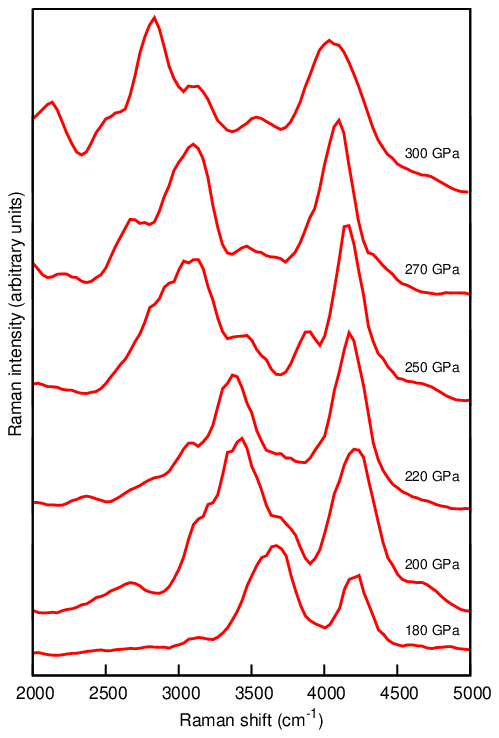}}}
\caption{\label{fig3} Calculated Raman signals from ab initio runs using CASTEP.
Simulations were equilibrated in Phase IV, then averages taken over 
3000 steps NVE pure H2, 288 atoms}
\end{center}
\end{figure}

\section{Raman results in Phase IV Hydrogen}
Phase IV in hydrogen has been identified as a layered structure (Fig
2) and the vibron spectrum shows one band of frequencies from each
layer. We have carried out extensive molecular dynamics using a density
functional theory code\cite{castep} to describe the electronic structure
and classical protons at a range of temperatures and pressures:
details were described previously\cite{Magdau}.

Fig 3 shows the variation with temperature and pressure.
The pressure evolution has been studied experimentally, and
the main features are all well reproduced by the calculation:
there are two strong peaks - indicating strongly bonded B layer and weakly bonded G layer;  There are sharp peaks in the B layer - indicating weak coupling and low anharmonicity; There are broad peaks in the G layer - indicating stronger intermolecular interactions which weaken the intramolecular bonds and increasing anharmonicity with pressure; there is a shoulder on the G-layer peak - indicating the high pressure onset of trimer rotation.

Temperature trends are less well studied experimentally, but from the 
simulations we predict two trends which should be measurable:
Softening of the B layer peak with increasing T; 
Broadening and splitting of the G - layer peaks.

It is important to note that the simulation extend far beyond the thermodynamic
stability  range of Phase IV.  We have seen evidence of spontaneous III-IV 
phase transitions in the MD, however this is strongly suppressed by the finite system size.

\section{Conclusions}
We have show that high temperature Raman data can be extracted from
molecular dynamics simulations without recourse to polarisability
calculations.  The projection method has been used widely before\cite{Pinsook,Souvatzis} for highly symmetric crystals where the
eigenmodes are determined by symmetry and the anharmonicity can be
used to renormalise their frequencies.   For 
hydrogen, where the normal modes themselves have only temporary validity, we have to extract the signal without recourse to normal modes.  
The basis of the method is the ansatz that the Raman signal can be associated directly with molecular motions (symmetric vibrons) and therefore calculated {\it without} explicit evaluation of phonon modes.   It must be admitted that the method is more challenging with respect to IR modes, which are related to out-of-phase vibrations: these are non-unique in a crystal with a large basis such as Phase IV hydrogen. 
However, some information can still be extracted from another ansatz - that the IR active modes are drawn from the high frequency end of each band and are of similar width to the Raman.


\begin{thebibliography}{12}
\bibitem{footnote}In fact, experimental H:D frequency ratios for
  vibrons are slightly lower than $\sqrt{2}$:1 in Phase
  IV, indicating that rotation is hindered and the 
  potential is slightly stiffer than harmonic.
\bibitem{footnote2}We did investigate placing large a energy in the
  vibron eigenmode and monitoring its decay.  This showed that the
  B-layer modes are longer-lived and the peaks should be sharper, but
  did not give quantative results.
\end{thebibliography}
\end{document}